\documentstyle[11pt,psfig]{article}
\begin{document}
\begin{center}
\Large{Circumstellar structure of RU Lupi down to au scales}

\vspace{6mm}
\large{M. Takami$^1$, J. Bailey$^2$, T.M. Gledhill$^1$, A. Chrysostomou$^1$ and J.H. Hough$^1$}
\end{center}

\vspace{4mm}
$^1$Department of Physical Sciences, University of
Hertfordshire, Hatfield, Herts AL10 9AB, UK. \\

$^2$Anglo-Australian Observatory, PO Box 296, Epping, NSW 1710, Australia \\

\vspace{4mm}
Accepted by MNRAS

\vspace{1cm}
\begin{center}
\large{\textbf{Abstract}}
\end{center}
\normalsize

We have used the technique of spectro-astrometry to study the milli-arcsecond
scale structure of the emission lines in the T Tauri star RU Lupi. The wings of
the H$\alpha$ emission are found to be displaced from the star towards the
south-west (blue wing) and north-east (red wing) with angular scales of 20-30
milli-arcsecs. This structure is consistent with a bipolar outflow from the
star.
From a study of the variability of the intensity and position spectra 
we argue that a combination of magnetically-driven bipolar outflow and
accreting gas contribute to the H$\alpha$ emission.
On the other hand,
the [OI] and [SII] emission are displaced from the star to the south-west
but at much larger distances than the H$\alpha$, hundreds of milli-arcsecs
for the high-velocity component (HVC) and down to 30 milli-arcsecs for
the low-velocity components (LVC). 
The presence of both red-shifted and blue-shifted outflows in H$\alpha$ but
only a blue-shifted outflow in the forbidden lines can be explained if  the
disc obscures the red-shifted forbidden line outflow, but a disc gap with outer
radius 3-4 au allows the red-shifted H$\alpha$ to be seen. This gap could be
induced by an unseen companion.

\twocolumn

\setlength{\parindent}{4mm}
\section{Introduction}
Optical emission lines from pre-main-sequence stars are
observed as clues to understanding the acceleration of jets and mass accretion
on to the stars. The spatial resolution of most techniques is
insufficient to resolve their dynamics directly, while
high-resolution spectroscopy has provided a variety of information
for these mechanisms. H$\alpha$ emission
shows a rich variety of profiles due to an interplay
between emission and absorption features (e.g., Reipurth, Pedrosa, \& Lago
1996), and various models have been discussed to explain their profiles
using accelerating stellar wind (e.g., Decampli 1981), decelerating stellar wind
(Mitskevich, Natta, \& Grinin 1993), disc wind (Calvet, Hartmann, \&
Hewett 1992), and infalling envelope (e.g., Calvet \& Hartmann 1992).
The forbidden lines often have two blue-shifted velocity components,
which have provided constraints on geometry and mechanism of
jet and/or wind (Eisl\"offel et al. 2000,
and references therein). The lack of red-shifted components
of these line supports the existence of circumstellar discs
(Eisl\"offel et al. 2000, and references therein).
In addition to high-resolution spectroscopy,
some techniques have been tried in the last decade
to obtain spatial information. 
Speckle observations reveal elongated structure of the
H$\alpha$ emission with an angular scale
of 0.09 arcsec towards T Tau (Devaney et al. 1995).
The $Hubble$ $Space$ $Telescope$ has provided fine structure of jets
towards young stellar objects (Ray et al. 1996; Burrows et al. 1996)
although the bandwidths of its filters are often insufficient to
suppress the stellar continuum from pre-main-sequence stars.
Ground-based long-slit spectroscopy is effective in
measuring the typical angular scale of a few forbidden lines 
suppressing the contamination of the strong continuum emission
(e.g., Hirth, Mundt, and Solf 1997).

Here we present the results of spectro-astrometric observations of RU Lupi.
The spectro-astrometry technique studies the small scale structure by observing
the relative position of the
centroid of the point-spread function on scales of a few milli-arcsec (Bailey
1998a, b).
The target, RU Lupi, is one of the most active T-Tauri stars in the
southern sky. The H$\alpha$ emission has broad wings up to 900 km s$^{-1}$
(Reipurth, Pedrosa, \& Lago 1996), and the equivalent width of the emission line is up to 200
{\AA}, which is one of the largest among T-Tauri star (Giovannelli et al. 1995).
Forbidden lines of [SII] and [OI] are also present in the spectrum of RU Lupi,
and at least the [OI] lines have an extended blue wing up to 300 km s$^{-1}$ and
a redder peak up to 30 km s$^{-1}$ (Gahm, Lago, \& Penston 1981; Hamann \&
Persson 1992). 
In this paper, the detail of the observations and the data reduction are described in
\S 2, the results are shown in \S 3, and the origin of H$\alpha$ emission and
the forbidden lines are discussed in \S 4 together with the existence of a
circumstellar disc with a gap. Throughout this paper, the distance of 140 pc to
the Lupus cloud is adopted based on Hughes, Hartigan, \& Clampitt (1993).

\section{Observations \& Data Reduction}
Observations were carried out on 1996 Aug 25, 1997 Jun 27, and 1999 Jul 2
at the 3.9-m Anglo-Australian Telescope using the RGO spectrograph.
The configuration with a 1-arcsec slit width, a 1200 line mm$^{-1}$ grating,
and the 82-cm camera provided a spectral resolution
($\lambda / \Delta \lambda$) of 7000.
The first two observations used a Tektronix 1024$\times$1024 thinned CCD.
The 1999 Jul 2 observation used a 2048$\times$4096 deep depletion CCD from
MIT Lincoln Labs.
The pixel scale was 0.23 arcsec with the Tektronix CCD and 0.15 arcsec with the MITLL CCD, which  provides
good sampling of the seeing profile (1--1.5 arcsec) at each wavelength.
The spectra were obtained at four slit position angles
(0$^{\circ}$, 90$^{\circ}$, 180$^{\circ}$, and 270$^{\circ}$).
The flat fields 
were made by combining
many exposures with the spectrograph illuminated by a tungsten lamp.
Wavelength calibrations were made by observations of
a CuAr lamp.

The data were reduced using the FIGARO package.
After subtracting the bias level and flat-fielding,
the position spectrum was determined by fitting the seeing profile
at each wavelength with a Gaussian function.
About 1-2$\times10^4$ photons at each wavelength provides
a typical accuracy of 2-5 milli-arcsec at the continuum level.
Any instrumental effect in the position spectra were eliminated
with high accuracy by subtracting those with opposite position angles
(0 $^{\circ} - 180^{\circ}$, or 90 $^{\circ} - 270^{\circ}$),
since these will remain constant on rotation whereas any
real structure in the source will reverse its sign. Effects which are
removed in this way include any curvature or optical distortion introduced
by the spectrograph, any misalignment of the spectrum with the CCD columns,
and any effects due to departure of the CCD pixels from a regular grid,
imperfect flat fielding, or charge transfer deficiencies in the CCD.
The derived position spectra have wavelength coverages of 6520-6750 {\AA}
for the former two runs and 6160-6780 A for the latest run,
which include H$\alpha$, [SII] 6716/6731 A, [NII] 6583 A, and [OI] 6300 {\AA}.
Each position spectrum has an arbitrary zero point, which
we adjust to correspond to the continuum position.
In addition to the position spectra, the intensity spectra were
obtained by subtracting the bias, flat-fielding, subtracting
the adjacent sky and extracting bright columns on the CCD. 
In all the spectra, the radial motion of the star is calibrated using
the Li 6707.815{\AA} absorption line on the stellar atmosphere.
The measured radial velocity in the local standard
rest frame is $+ 8 \pm 2$ km s$^{-1}$.

Similar methods were previously used by Hirth, Mundt, and Solf (1994, 1997)
to study forbidden emission lines in pre-main sequence stars. However, the
techniques described here enable us to achieve substantially higher accuracy
in the relative positions. 

\section{Results}

\subsection{Positional displacement and intensity profile of H$\alpha$,
[OI] and [SII] emission}

Fig. 1 shows the raw spectra of the intensity and the positional displacement.
In the intensity spectra, a number of emission lines including
H$\alpha$, HeI, [SII], [OI], [NII], FeI, FeII, and SiII are detected.
The wavelengths and equivalent widths without atmospheric correction
are summarized in Table 1.

Among these lines, the H$\alpha$, [SII] 6731/6716{\AA},
[NII] 6583 {\AA}, and [OI] 6300 {\AA} lines show displacements from the continuum
position in the position spectra. 
The positions shown in Fig. 1 are the centroid of the combined line and 
continuum emission of the star, that is, the centroid of
their positions weighted by their intensities.
To get the true position of the emission
line region the observed position must be corrected for the effect of the
underlying continuum by dividing the positional displacement in the raw spectra
by $I_{\lambda \rm{(line)}}/(I_{\lambda \rm{(line)}}+I_{\lambda \rm{(conn)}})$.
Fig. 2 shows the positional displacement of
H$\alpha$, [OI] 6300 {\AA}, and [SII] 6731/6716 {\AA} observed on 1999 Jul 2
after this calibration.
Typical uncertainties for the displacements in Fig. 2 are
2, 15, 40, and 100 milli-arcseconds for H$\alpha$, [OI] 6300 {\AA}, [SII] 6731 {\AA} and 6716 {\AA},
respectively. For H$\alpha$, higher accuracy was obtained
than that at the continuum level because of its brightness while the accuracy
of the other forbidden lines are much less than the continuum
because of their lower intensity. The line profiles shown in Fig. 2 are similar
to those previously presented by Giovannelli et al. (1995), Reipurth,
Pedrosa, \& Lago (1996), and Hamann (1994).

Fig. 2 shows that the positional displacement of H$\alpha$ emission extends
towards the south-west and the north-east over angular scales of up to 20-30
milli-arcsec.
The intensity profile of H$\alpha$ emission is symmetric except
for a redshifted absorption dip at +50 km s$^{-1}$ from the star, and the
blue-shifted wing corresponds to the displacement at the south-west  while the
red-shifted wing corresponds to that at the north-east. At these wings, the
displacement monotonically increases with the relative velocity from the star
in the velocity ranges of $-50$ to $-200$ km s$^{-1}$ and $+200$ to $+300$ km
s$^{-1}$. On the other hand, the displacement is distributed within 5
milli-arcsec from the star in a lower velocity range between $-50$ and $+200$
km s$^{-1}$, and does not show clear correlation with the velocity in Fig. 2.

The positional displacement of the three forbidden lines
extends only towards the south-west, the same direction as those
of the blue-shifted H$\alpha$ wing, and that of HH55 which lies 3 arcmin
from RU Lupi (cf. Krautter, Reipurth, \& Eichendorf 1984).
The displacements are larger than that of H$\alpha$ emission
by a factor of 10-30, and smallest in [OI] and 
largest in [SII] 6716{\AA}: up to 200, 400, and 600 milli-arcsecs
for [OI] 6300{\AA}, [SII] 6731 {\AA}, and 6716{\AA}, respectively.
These line profiles show two peaks
with typical velocities between $-250$ and $-100$ km s$^{-1}$ for the high
velocity component (HVC) between $-100$ and $0$ km s$^{-1}$ for the low velocity
component (LVC). The HVC is more extended spatially
than in the LVC as in other T Tauri stars.
The typical angular scales of the LVC are
30-40 milli-arcsec in [OI] 6300 {\AA} and 150 milli-arcsecs in [SII] 6731 {\AA}.
Fig. 2 does not show a clear difference in
the displacement between the two components in [SII] 6716{\AA}.

The relation between the velocity and the radial distance of the
displacement
for H$\alpha$, [OI] 6300{\AA}, and [SII] 6731{\AA},
is plotted in Fig. 3.
A higher signal to noise ratio of the positional displacements
has been obtained by binning along the wavelength.
This figure shows that the velocity monotonically increases
with the distance for all the lines, and the gradient for H$\alpha$ emission
is much larger than those for the forbidden lines.
Fig. 3 also shows slight asymmetry of the velocity field between
the blue-shifted and red-shifted wings of H$\alpha$ emission:
the positional displacement is prominent in the figure beyond the velocity of
100 km s$^{-1}$ for the blue-shifted wing and 200 km s$^{-1}$ for
the red-shifted wing.
In the forbidden lines, the displayed velocity gradient is larger
in the HVC ($\Delta$V $>$ 100 km s$^{-1}$) and smaller in the LVC
($\Delta$V$ < $100 km s$^{-1}$). 
On the other hand, the LVC in [OI] reveals no velocity
gradient while those in the [SII] 6731 {\AA} line show slight positive
gradients.
The position of the LVC is offset from the star even at the zero velocity:
30 and 70 milli-arcsecs for [OI] and [SII] 6731 {\AA}, respectively.

The observed tendencies for the line profiles and the positional
displacement of the HVC agree with
the previous observations of other T Tauri stars.
Ground-based long slit spectroscopy shows the positional
displacement of the HVCs of T Tauri stars in the Taurus-Auriga cloud
(Hirth, Mundt, and Solf 1997),
which lies at a similar distance (140 pc) to that of the Lupus cloud.
The HVC and LVC in these objects have typical velocities
$-50$ to $-150$ km s$^{-1}$ and $-5$ to $-20$ km s$^{-1}$, respectively,
which are similar to those of RU Lupi.
The measured displacements of the HVC are 0.2 and 0.6 arcsecs for [OI] and [SII] 6731{\AA},
respectively, which are similar to those seen in our observations.
The [SII] lines have much lower critical density than the [OI] line
(typically 10$^3$ and 10$^6$ cm$^{-3}$, respectively), and the observed
positional difference between these lines indicates that the outer region
has a lower electron density. Such a distribution of the electron density
could be due to diverging stream lines of the outflow,
as suggested by Hirth, Mundt, and Solf (1997).

\subsection{Time variation of H$\alpha$ and [SII] emission}

Fig. 4 shows the time variation of the intensity profile and the positional
displacement for H$\alpha$, [SII] 6731{\AA}, and [SII] 6716{\AA} emission. Since
photometric observations were not made on the same dates, each line profile
is displayed in a unit of the continuum intensity on each date. The positional
displacements are displayed after removal of the contamination of the continuum, as in Fig. 2.

In the case of H$\alpha$, time variation was detected for both the line profile
and the position spectra. The intensity profiles
observed on 1996 Aug 25 and 1997 Jun 27 are nearly the same at the wings. On the other hand, at the peak with the velocity
of $-100$ to $+100$ km s$^{-1}$, the profile on the latter date
has lower intensity relative to the continuum than that on the former date.
On 1999 Jul 2, the relative intensity of the H$\alpha$ to the continuum
is slightly lower than the others except that at the absorption dip.
In the same observing run, the absorption dip is shallower
and relative intensity is higher at the top than those observed in the
other runs.
Clear change of the position spectra of H$\alpha$ was observed
during 1997-1999 with the velocity ranges
of $-200$ to $-100$ km s$^{-1}$ and $+100$ to $+300$ km s$^{-1}$.
The positional displacements of these
wings are larger on the latest date than
in the others by a factor of 2-3.
The repeatability of the measured positional displacement
between $-100$ and $+100$ km s$^{-1}$ confirms
that the displacement is not due to slight oscillation
of the tracking of the telescope, since any other pre-main sequence
stars observed in 1999 do not show the same displacements as observed
in this object.

The relative intensity of the [SII] lines to the continuum is
smallest on the earliest date, and largest on the latest date
at the whole velocity range. 
On the other hand, the peak velocity and the line widths of the HVC and LVC
did not change throughout the observations. 
To determine the difference of the time variation between the HVC and LVC,
we measure the equivalent widths
of the HVC and LVC respectively, and these are tabulated in Table 2.
For both the [SII] 6716{\AA} and 6731{\AA}, the LVC shows a larger increase in
the equivalent width than the HVC: the difference of the relative
intensities between 1996 and
1999 is 40-50 and 10-30 percent for the LVC and HVC, respectively.
Table 2 also shows the equivalent width ratio of the two lines
of 0.53-0.60 and 0.37-0.40 for the HVC and LVC, respectively,
and these ratios indicate electron densities of 4-6$\times10^3$ and more than
$10^4$-$10^5$ cm$^{-3}$ assuming a electron temperature of $10^4$ K
(cf. Osterbrock 1989). Such a difference of the electron density
between the HVC
and LVC was also observed in other T Tauri stars by Hamann (1994).
On the other hand, Fig 4. shows no systematic time variation
in the position spectra of the two [SII] lines.
The displayed variation would indicate motion of gas, although
higher signal-to-noise ratio is required to comfirm the existence
of the variation.

In addition to the time variation of the emission line fluxes,
variation of the continuum flux can affect the intensity profiles shown
in Fig. 4. Its contribution is not clear owing to the lack
of photometric data. Previous observations show the time variation
of the continuum which may be caused by obscuration by clumpy clouds
(Gahm et al. 1974;
Giovannelli et al. 1995). In addition, Table 1 shows time variation
of the equivalent width for the Li absorption, which suggests
variation of the veiling continuum.
The intensity profiles of H$\alpha$ and the two [SII] lines
show different time variation, and it cannot be simply explained
by variation of the continuum. The ratios of the equivalent width
between the LVC and the Li absorption are nearly constant over time
(1.4-1.6 and 0.5-0.6 for [SII]6731{\AA}/Li and [SII]6716{\AA}/Li, respectively),
and suggest that the time variation of the LVC is due to the variation
of the veiling continuum.

\subsection{Comparison with the other lines profiles}

The profiles of the other lines are plotted in Fig. 5.
Most of the permitted lines including FeI, FeII, SiII have triangular profiles with
peak velocities of $-50$ to $+50$ km s$^{-1}$, and full width
zero intensities of less than 250 km s$^{-1}$.
The width of these lines is much narrower than that of the H$\alpha$, and
the velocity ranges of these lines do not overlap that of the H$\alpha$ wings:
thus, these lines are not related to the H$\alpha$ wings observed in 1999.
On the other hand, the profiles of [OI] 6364 {\AA} and [NII] 6583 {\AA} are much more
similar to those of [OI] 6300 {\AA} and the two [SII] lines than that of H$\alpha$.
The profiles of [OI] 6364 {\AA}, which shares the upper transition level
with [OI] 6300 {\AA}, is nearly the same as that of the latter line shown
in Fig. 2. No clear signature of the LVC is found in the [NII] profile
as in the case of other T Tauri stars (e.g., Hartigan, Edwards, and Ghandour 1995).

The HeI 6678 {\AA} line shows different profiles from the others, and
a prominent time variation was observed.
In 1996 and 1997, the line had nearly the same triangular profile
with a full width zero intensity of about 400 km s$^{-1}$.
On the other hand, in 1999, the line width was reduced to be about a half
without changing the peak velocity and the relative intensity to the continuum at the peak.
Two components with different line widths exist in the latter profile: a narrow component
(NC) with a full width half maximum of about 50 km s$^{-1}$, and a broad
component (BC) with a full width zero intensity of more than 200 km s$^{-1}$.
Such coexistence of two components has been often observed 
in HeI 5876{\AA}, CaII, and FeII lines of other T Tauri stars (e.g.,
Batalha et al. 1996, Muzerolle, Hartmann, \& Calvet 1998, Beristain, Edwards,
\& Kwan 1998), and both the components are considered to be associated with mass
accretion in the vicinity of the star (e.g., Najita et al. 2000, and
references therein). As shown in Fig. 4, H$\alpha$ profiles have
a red absorption dip indicative of infalling material to the star.
The H$\alpha$ and HeI profiles observing in 1999 are
characterized by a shallower redshifted absorption dip
and a weaker broad component, respectively,
and such relation between the H$\alpha$ dip
and the HeI broad component suggests that the latter component
arises from gas accreting to the star.
No prominent emission associated with
the H$\alpha$ wings was found in the broad component of HeI in 1999:
thus, we conclude that the H$\alpha$ wings have a different origin
from that of the broad component of the HeI line.

\section{Discussion}
\subsection{Origin of H$\alpha$ emission}
Positional displacement as measured by spectro-astrometry should be sensitive
to any asymmetric structure about the continuum source.
In the case of H$\alpha$ emission from pre-main-sequence stars,
binary companions are responsible for the displacement
as shown by Bailey (1998a).
However, the feature seen in RU Lupi is quite unlike
those normally seen in binary systems which are not symmetric about the line
centre. Binary stars with the same spectrum and different radial velocities 
could produce the observed positional displacement, although such motion at the
observed direction does not agree with the orbital motion perpendicular to the
outflow. No binary companion has been detected by $the$ $Hubble$ $Space$
$Telescope$ (Bernacca et al. 1995) or infrared speckle observations (Ghez et al.
1997). Thus, it is likely that the detected feature reflects the distribution
of circumstellar matter around the star.

A variety of mechanisms have been previously proposed to explain the
observed H$\alpha$ profiles from pre-main-sequence stars.
Magnetically-driven stellar winds
which accelerate with distance have been suggested for
 Balmer emission including H$\alpha$ (e.g., Decampli 1981; Hartmann,
Edwards, \& Avrett 1982; Lago 1984; Natta, Giovanardi, \& Palla 1988),
while a decelerating "stochastic" stellar wind was proposed by Mitskevich,
Natta, \& Grinin (1993). Calvet \& Hartmann (1992) and Hartmann,
Hewett \& Calvet (1994) argue that the H$\alpha$ emission is produced in
infalling envelopes. Edwards et al. (1994) presented for 15 T Tauri stars a set
of high resolution profiles including Balmer, HeI, and NaI lines and showed that the
blue/red asymmetry of upper Balmer lines agrees well with magnetospheric
accretion models. On the other hand, Reipurth, Pedrosa, and Lago (1996)
compared their H$\alpha$ profiles with various models, and suggested that a single model
cannot account for the variety of the H$\alpha$ profiles. 
In addition to outflowing and inflowing motion, Keplerian rotation would also be
responsible for the shape of the H$\alpha$ profile.

Among these candidates, we consider bipolar outflow as a plausible explanation
for the observed displacement for the following reasons:
(1) the positional displacement of the blue-shifted component
is in the same direction as those of forbidden lines,
which are considered to trace outflow, and
(2) the other permitted lines, which probe mass accretion, do not
have such broad wings as H$\alpha$, thus these wings should have
different origin (cf. \S 3-3).
No clear evidence for inflowing motion has been detected in the
position spectra, consistent
with previous models which propose that the H$\alpha$ emission
arises from a region within less than ten stellar radii
(e.g., Calvet \& Hartmann 1992; Hartmann, Hewett \& Calvet 1994).
Rotation cannot explain the
observed displacement for the following reasons:
(1) the measured velocity increases with positional displacement
while Keplerian rotation should behave in an opposite fashion, and
(2) the star needs to be more than 100 M$_{\odot}$ to bound the gas
with the measured positional displacement and velocity (3 au and 200 km s$^{-1}$,
respectively) while previous research suggests the mass of the star is
0.5-2 M$_{\odot}$ (Gahm et al. 1974; Lamzin et al. 1996).

Figs 2 and 3 show that the velocity of the outflowing gas increases with
positional displacement, a tendency which agrees with
that of magnetically-driven wind models which predict that
the flow accelerates with distance (e.g., Decampli 1981; Hartmann,
Edwards, \& Avrett 1982; Lago 1984). 
Most of the models adopt a spherically symmetric line forming region
not exceeding ten stellar radii, and predict that the
acceleration of the flow ends within this scale.
On the other hand, the observed position spectra of RU Lupi
indicate bipolar geometry and a velocity gradient extending
to a few au (Figs 2 and 3).
The latter results would indicate slower acceleration of the
outflow than those predicted by models,
however, models with more appropriate geometry and
direct prediction of the positional displacement
are necessary for detailed comparison.
Actually, H$\alpha$ emission from pre-main-sequence stars should be
optically thick (cf. Reipurth, Pedrosa, \& Lago 1996),
thus the observed velocity field would be highly sensitive to the geometry
which affects radiative transfer.

The time variation of the positional displacement and the intensity profile
for the H$\alpha$ wings are similar to that of HeI in the following points: there is
no clear difference between 1996 and 1997, and a remarkable difference
between 1997 and 1999. 
Since the HeI emission is related to the accreting gas,
as shown in \S 3-3, such similar time variation suggests
that the accreting gas also contributes to the H$\alpha$ flux
at least in 1996 and 1997.
Even if the positional displacement of the bipolar outflow
is constant over time, the net positional displacement can change
as the flux of the accreting gas at the centre varies.
Transient time variation of the displacement in high-velocity
wings could be explained by direct motion of the outflowing ionized gas,
although we reject this possibility since line profiles
with the same velocities only show slight change of line-to-continuum ratio
at the same velocities.

Here we attempt to explain the observed positional displacement
with a steady bipolar outflow and time-variable flux from inflow.
To remove contamination from inflow on the total H$\alpha$
flux, and reproduce the position spectra of the outflow
we make the the following assumptions:
(1) the H$\alpha$ emission from infalling gas
arises from the star and cannot be spatially resolved
by our observations,
(2) inflow contributes
to the observed H$\alpha$ profiles only in 1996 and 1997,
and (3) the time variation of the
Li equivalent widths shown in Table 1
is due to the veiling continuum. The first assumption provides
the observed net positional displacement as follows:
\begin{equation}
\Delta x_\lambda
= \Delta x_{\lambda \rm{out}}
     \cdot \frac{I_{\lambda \rm{(out)}}}
                {I_{\lambda \rm{(out)}}+I_{\lambda \rm{(in)}}}              
\end{equation}
where $\Delta x_\lambda$ is the net displacement shown in Fig. 4,
$\Delta x_{\lambda \rm{out}}$ is the positional displacement caused by
the bipolar outflow, 
$I_{\lambda \rm{(out)}}$ and $I_{\lambda \rm{(in)}}$,
are the absolute H$\alpha$ intensity
from the outflow and inflow, respectively.
This assumption is consistent with our observations, and the previous
inflow models described above.
From the second assumption, the intensities from the outflow and inflow 
are obtained as follows:
\begin{equation}
I_{\lambda \rm{(out)}} = I_{\lambda \rm{(H \alpha 99)}}, \hspace{0.5cm}
I_{\lambda \rm{(in)}}  = I_{\lambda \rm{(H \alpha )}}
                           - I_{\lambda \rm{(H \alpha 99)}},
\end{equation}
where $I_{\lambda \rm{(H \alpha )}}$ is the absolute H$\alpha$ intensity
in a certain observing run,
and $I_{\lambda \rm{(H \alpha 99)}}$ is that observed in 1999.
This assumption is plausible for the H$\alpha$ wings
since those observed in 1999 are not associated with the other
permitted lines indicative of mass accretion as described in 3-3.
The third assumption provides the absolute H$\alpha$ intensities
from the observed parameters as follows:
\begin{equation}
\frac{I_{\lambda \rm{(H \alpha )}}}{I_{\lambda \rm{(H \alpha 99)}}}
 = \frac{i_{\lambda \rm{(H \alpha )}} \cdot I_{\lambda \rm{(conn)}}}
        {i_{\lambda \rm{(H \alpha 99)}}  \cdot I_{\lambda \rm{(conn99)}}} \nonumber
 = \frac{i_{\lambda \rm{(H \alpha )}}}{i_{\lambda \rm{(H \alpha )}}} \cdot
   \frac{W_{\lambda \rm{(Li99)}}}{W_{\lambda \rm{(Li)}}},
\end{equation}
where $i_{\lambda \rm{(H \alpha )}}$ and $I_{\lambda \rm{(conn)}}$ are the relative
H$\alpha$ intensity to the continuum and the absolute intensity of the continuum
in a certain observing run, respectively, and
$i_{\lambda \rm{(H \alpha 99)}}$ and $I_{\lambda \rm{(conn99)}}$ are those
observed in 1999.
$W_{\lambda \rm{(Li)}}$ and $W_{\lambda \rm{(Li99)}}$ are the equivalent widths
of the Li absorption in a certain observing run and in 1999, respectively. 
By combining equations (1)-(3),
the positional displacement of the outflow is obtained
from the observed parameters as follows:
\begin{equation}
\Delta x_{\lambda \rm{(out)}}= \Delta x_\lambda \cdot 
   \frac{i_{\lambda \rm{(H \alpha )}}}{i_{\lambda \rm{(H \alpha 99)}}} \cdot
   \frac{W_{\lambda \rm{(Li99)}}}{W_{\lambda \rm{(Li)}}},
\end{equation}

Fig. 6 shows the reproduced position spectra using equation (4).
The spectra for 1999 are identical with the observed one
shown in Fig. 4, since we assume that the outflow dominates the observed
intensity in this observing run.
These spectra agree well with each other, thus support the idea that the time
variation of
the positional displacement can be explained by the combination of steady
bipolar outflow and
time-variable flux from the accreting gas.

The contribution of the inflow to the total H$\alpha$ flux could be
estimated from equations (2) and (3): according to these equations,
32 and 28 percents of the total H$\alpha$ flux would arise from
the accreting gas in 1996 and 1997, respectively.
These values would be lower limits,
since the inflow with low velocities would contribute to
the H$\alpha$ emission in 1999.
The permitted lines related to mass accretion
were detected with a typical velocity range between $-150$ and
$+150$ km s$^{-1}$ as shown in \S 3-3, and there is no reason to assume
that H$\alpha$ emission
does not arises from the same region.

\subsection{Origin of the forbidden lines}
Optical forbidden emission lines including [OI] 6300{\AA} and [SII] 6716/6731{\AA}
provide information on the outflows from T Tauri stars.
These line profiles often have two blue-shifted velocity components:
a high-velocity component (HVC) and low-velocity component (LVC) which
have typical velocities of $-$50 to $-$150 km s$^{-1}$ and $-5$ to
$-20$ km s$^{-1}$, respectively (Hirth, Mundt, and Solf 1997).
The HVC is spatially resolved by ground-based long-slit
spectroscopy in  the nearest star-forming regions
(e.g., Solf \& B\"ohm 1993; Hirth, Mundt, and Solf 1997).
Hirth, Mundt, and Solf (1997) have
shown that the typical angular scale of the HVC is 0.2--0.6 arcsec from their
large sample of data mainly in the Taurus-Auriga cloud.
The high speed and narrow linewidth
of the HVC indicates that the velocity vectors of the contributing
gas particles are nearly parallel to one another, i.e.,
suggesting a well-collimated flow or jet (Kwan \& Tademaru 1988).
On the other hand, the low-velocity component (LVC) has smaller angular scales 
in the same sample (0.1-0.2 arcsec or less), and its origin is less
understood.
Geometric effects of a single outflow are proposed to explain
the existence of the two velocity components (e.g., Edwards et al. 1987;
Hartmann \& Raymond 1989), while two distinct mechanisms,
jet and disc wind, are proposed for each velocity component
by Kwan and Tademaru (1988, 1995).
Hamann (1994) measured the electron density, temperature,
and the ionization state between the HVC and LVC in many T Tauri stars,
and the measured difference of the physical conditions
 suggest distinct origins for these two components as
proposed by Kwan and Tademaru (1988, 1995). Similar results were
also presented by Hartigan, Edwards, \& Ghandour (1995).
The velocity dispersion and flow velocity are comparable in the LVC,
suggesting moderate
collimation such as that due to a disc wind (e.g., Solf \& B\"ohm 1993).
Most pre-main-sequence stars do not have red-shifted
components to their forbidden lines, and this is explained by
obscuration of these components by circumstellar discs (e.g., Eisl\"offel et al. 2000,
and references therein).

The [OI] and [SII] lines from RU Lupi have two blue-shifted components, and our results also support
the existence of two distinct mechanisms.
The [SII] intensity ratio indicates different electron densities
between the HVC and LVC as shown in \S 3-2 (4-6$\times10^3$ and more than
$10^4$-$10^5$ cm$^{-3}$ assuming a electron temperature of $10^4$ K),
a tendency also observed
in other T Tauri stars by Hamann (1994). In addition, the intensity of each
component exhibits different time variations as described in \S 3-2.
The coexistence of two distinct mechanisms can easily explain
these differences.

Furthermore, the velocity field in the LVC shown in Fig. 3
agrees with the disc wind hypothesis.
Fig. 3 shows that the position of the LVC is offset from
the star even at zero velocity and
nearly constant over all velocities in [OI] emission,
as described in \S 3-2.
The former cannot be explained
by acceleration of outflow, while
both tendencies can easily be explained by moderate collimation and/or
rotation which would be expected in a disc wind.
In the case of [SII] 6731{\AA}, the positional
displacement in the LVC increases with the velocity. Such
discrepancy between the [OI] and [SII] emission
can be explained by blending of the HVC and LVC near
their boundary.
The profiles of the forbidden lines in Fig. 2
show that the two components are partially
blended, thus contamination from the HVC could increase
the positional displacement in the LVC near their boundary.
It is likely that this effect is more prominent for the [SII]
line, since the lower critical density of this line
allows for larger contamination from the HVC which has a lower
electron density (cf. \S 3-2).

\subsection{Circumstellar disc and disc gap}
As shown in Fig. 2, the forbidden lines have only blue-shifted
components, similar to many other T Tauri stars, a fact
usually explained by obscuration of the red-shifted flow by
a circumstellar disc (e.g., Eisl\"offel et al. 2000, and 
references therein). Infrared excess and a bright millimetre continuum
suggest the existence of a circumstellar disc around RU Lupi
(e.g., Giovannelli et al. 1995; Carballo et al. 1992;
N\"urnberger et al. 1997), supporting
the explanation for the lack of red-shifted components.

On the other hand, H$\alpha$ emission has both blue- and red-shifted
components. The measured positional displacement for the H$\alpha$ emission is
smaller than any velocity component of [OI] and [SII] emission.
The detection of the red-shifted H$\alpha$ outflow can be explained
by a gap in the disc which allows this emission to be seen.
The presence of disc holes and gaps is often inferred around other
pre-main-sequence stars by infrared spectral energy
distributions (SEDs) (e.g., Mathieu et al. 1991; Marsh \& Mahoney 1992; Jensen
\& Mathieu 1997), and directly observed in near-infrared imaging
(e.g., Roddier et al. 1996; Silber et al. 2000) and at radio
wavelengths (e.g., Dutrey, Guilloteau, \& Simon 1994).
Fig. 3 suggests that the projected radius of the hole or gap is
20-30 milli-arcsec, which corresponds to 3-4 au, assuming
symmetric distribution
of the blue-shifted and unseen red-shifted components
of the forbidden lines.

To investigate the presence of a disc gap, we plot the infrared SED in Fig. 7. The flux at $JHKLM$-bands and
12, 25, 60, 100 $\mu$m were obtained from Giovannelli et al. (1995) and
Carballo et al. (1992).
Giovannelli et al. (1995) show the near-infrared flux is highly variable with time,
thus the data-set with the largest and smallest flux were selected from their data
and plotted in the figure.
In Fig. 7, a spectrum obtained
via the $Infrared$ $Space$ $Observatory$ (ISO) data archive is
also plotted to interpolate the flux in the mid-infrared.
Lamzin et al. (1996) claim that the mid-to-far infrared flux
obtained by Carballo et al. (1992) would be overestimated,
since another source discovered by Cohen \& Schwartz (1987)
is included in the same aperture.
However, the consistency of their data with the ISO spectrum
shows that the contribution from this source is much less than
the total flux: the aperture of the ISO Short-Wave Spectrometer (SWS)
with a size of 20$\times$33 arcsec$^2$ does not include
this source which lies about 3 arcmin away from the target.

Fig. 7 shows two peaks in the SED which arises from dust.
The near-infrared peak at 1.65 $\mu$m is longer
than the peak for the star (1.0 $\mu$m) with an effective temperature
of 3900 K (N\"urnberger et al. 1997; Lamzin et al. 1996).
Previous authors show that the selective extinction towards RU Lupi is
0.3-1.3 A$_V$ (Hughs et al. 1994; Giovannelli et al. 1995).
Such values can shift the peak of the stellar continuum by 
less than 0.2 $\mu$m suggesting that the
near-infrared excess from the stellar photosphere represents
hot dust from the accretion disk. Such a contribution by the accretion disk
to the near-infrared SED is also supported by the
near-infrared colours (J-H and H-K of 0.84-0.92 and 0.72-0.77, respectively)
and diagrams shown by Hughes et al. (1994), Greene \& Meyer (1995), and
Itoh et al. (1999).
On the other hand, the mid-to-far infrared SED shows another
excess which is due to cooler dust with a temperature of less than 100 K.
The shallow dip at 4-15 $\mu$m suggests a lack of dusty
material at temperatures of 200-900 K. These temperatures
correspond to the radiative temperature at a radius of 0.1-2 au
from the star, assuming the total luminosity from
the star and accreting gas of 5 L$_{\odot}$ based on Lamzin et al. (1996).
These spatial scales agree well with that of the suggested gap
in Figs 2 and 3.

Theoretical work has shown that disc gaps and holes could be induced by
binary companions or young planets (e.g., Lubow \& Artimowicz 2000).
In the case of RU Lupi, no other pre-main-sequence companions
have been detected by $Hubble$ $Space$ $Telescope$ (Bernacca et al.
1995), infrared speckle observations (Ghez et al. 1997), or
spectro-astrometry which is sensitive to the presence of
binaries (cf. Bailey 1998a).
Therefore, the observed disc gap suggests the existence of
an unseen companion. A young planet could be one of possible
candidates for the following reasons: the presence of the fragmented dusty clouds around RU Lupi are suggested by variability at optical to infrared wavelengths (Gahm et al. 1974; Giovannelli et al.1995), and Gahm et al. (1974) claim that such clouds would indicate ongoing planetary formation around this object.
In addition, the observed disc gap has a similar
size to that of the orbit of the Jovian planets in our solar system,
and Takeuchi, Miyama, \& Lin (1996) suggest
that a proto-Jupiter could induce a disc gap with a similar size to its orbit.
Their numerical calculations show that a Saturnian and Jovian mass protoplanet
can open a gap in 10$^3$-10$^4$ and 10$^2$-10$^3$ yrs, respectively, timescales
which are much shorter than the age of RU Lupi (10$^6$ yrs - Lamzin et al. 1996).

\section{Conclusions}
Spectro-astrometric observations have been used to determine the
spatial distribution of optical emission lines from RU Lupi.
Positional displacements
were detected in H$\alpha$, [OI] 6300 {\AA}, [SII] 6731/6716 {\AA},
and [NII] 6583 {\AA} with accuracies down to a few milli-arcsec.
The intensity profiles were also obtained for these lines
together with HeI, FeI, FeII, SiII, and another [OI] line.

The positional displacement of H$\alpha$ emission extends
towards the south-west and north-east with angular scales
of 20-30 milli-arcsecs. The blue-shifted wing in the
intensity profile corresponds to the displacement
at the south-west while the red-shifted wing corresponds
to that at the north-east. We conclude that
the positional displacement of the H$\alpha$ emission
is due to bipolar outflow for the following reasons:
(1) the distribution of the positional displacement is
    almost symmetric between the blue-shifted and red-shifted
    components,
(2) the displacement of the blue wing aligns well with 
    that of the forbidden lines which are considered to
    trace outflowing gas.
The velocity of the outflow increases with the distance,
suggesting the flow is magnetically-driven.
On the other hand, the time variation of the intensity
profile and the positional displacement can be explained
well by the variation of H$\alpha$ flux from the accreting gas
at the star.
We estimate the contribution of the accreting gas to the
total flux to be more than 32 and 28 percent in 1996 and 1997,
respectively.

The positional displacement of [OI] and [SII] emission
extends towards the south-west, in the same direction
as HH55 which lies 3' from the star.
The intensity profiles have two blue-shifted components,
as seen in many other pre-main-sequence stars,
and the positional displacements of hundreds of milli-arcsecs
and down to 30 milli-arcsecs were detected for the
high-velocity components (HVC) and the low-velocity components
(LVC), respectively. The differences of the derived
electron density and time variation between the two components
suggest that these components have distinct origins,
e.g., jet and disc wind. The positional displacement in the LVC
is consistent with
the LVC originating from a disc wind.

The forbidden lines have only blue-shifted components,
as do those in many other T Tauri stars, and this is usually explained
by the obscuration of the red-shifted flow by a circumstellar disc.
On the other hand, the H$\alpha$ outflow has both blue-shifted
and red-shifted components. Such a difference suggests
the presence of a disc gap with an outer radius of 3-4 au.
The infrared spectral energy distribution is consistent with
the presence of a gap on this scale.
This gap could be induced by a unseen companion
such as a young planet. \vspace{1cm}

We would like to thank the staff at the Anglo-Australian Observatory
for their help and support during the observations.
We also thank ISO staff for their assistance
in using their archive system, and the referee (T.P. Ray)
for his valuable comments.
MT thanks PPARC for support through a PDRA.

\vspace{1cm}

\large{\textbf{References}}\textmd{}\normalsize
\setlength {\parskip} {2mm}
\setlength {\parindent} {0mm}

Bailey J., 1998a, MNRAS, 301, 161 \\
Bailey J., 1998b, SPIE Proceeing, 3355, 932 \\
Batalha C.C., Stout-Batalha N.M., Basri G., Terra M.A.O., 1996, ApJS, 103, 211 \\
Beristain G., Edwards S., Kwan J., 1998, ApJ, 499, 828 \\
Bernacca P.L., Lattanzi M.G., Porro, I., Neuh\"auser R., Bucciarelli B., 1995, A\&A, 299, 933 \\
Burrows C.J., Stapelfeldt K.R., Watson A.M., Krist J.E., Ballester G.E., Clarke J.T., Crisp D., Gallagher J.S.III, Griffiths R.E., Hester J.J., Hoessel J.G., Holtzman J.A., Mould J.R., Scowen P.A., Trauger J.T., Westphal J.A. 1996, ApJ, 473, 437Calvet N., Hartmann L., 1992, ApJ, 386, 239 \\
Calvet N.,Hartmann L., Hewett R., 1992, ApJ, 386, 229 \\
Carballo R., Wesselius P.R., Whittet D.C.B., 1992, A\&A, 262, 106 \\
Cohen M., Schwartz R.D., 1987, ApJ, 316, 311 \\
Decampli W.M., 1981, ApJ, 244, 124  \\
Devaney M.N., Thi\'{e}baut E., Foy R., Blazit A., Bonneau D., Bouvier J., de Batz B., Thom Ch., 1995, A\&A, 300, 181 \\
Dutrey A., Guilloteau S., Simon M., 1994, A\&A, 286, 149 \\
Edwards S., Cabrit S., Strom S.E., Heyer I., Strom K.M., Anderson E., 1987, ApJ, 321, 473 \\
Edwards S., Hartigan P., Ghandour L., Andruis C., 1994, AJ, 108, 1056 \\
Eisl\"offel J., Mundt R., Ray T.P., Rodr\'iguez L.F, 2000, Protostars and Planets IV, 815 \\
Gahm G.F., Nordth H.L., Olofsson S.G., Carlborg N.C.J., 1974, A\&A, 33, 399 \\
Gahm G.F., Lago M.T.V.T., and Penston M.V., 1981, MNRAS, 195, 59 \\
Ghez, A.M. McCarthy D.W., Patience J.L., Beck T.L., 1997, ApJ, 481, 378 \\
Giovannelli F., Vittone A.A., Rossi C., Errico L., Bisnovatyi-Kogan G.S., Kurt V.G., Lamzin S.A., Larionov M., Sheffer E.K., Sidorenkov V.N. 1995, 114, 341Greene T.P., Meyer M.R., 1995, ApJ, 450, 233 \\
Hamann F., 1994, ApJS, 93, 485 \\
Hamann F. Persson S.E., 1992, ApJS, 82, 247 \\
Hartigan P., Edwards S., Ghandour L., 1995, ApJ, 452, 736 \\
Hartmann L., Hewett R., Calvet N., 1994, ApJ, 426, 669 \\
Hartmann L., Edwards S., Avrett E., 1982, ApJ, 261, 279 \\
Hartmann L. Raymond J.C., 1989, ApJ, 337, 903 \\
Hirth G.A., Mundt R., Solf J., 1994, A\&A, 285, 929 \\
Hirth G.A., Mundt R., Solf J., 1997, A\&AS, 126, 437 \\
Hughes J., Hartigan P., Clampitt L., 1993, AJ, 105, 571 \\
Hughes J., Hartigan P., Krautter J., Kelemen J., 1994, AJ, 108, 1071 \\
Itoh Y., Tamura M., Nakajima T., 1999, AJ, 117, 1471 \\
Jensen E.L.N. Mathieu R.D., 1997, AJ, 114, 301 \\
Krautter J., Reipurth B., Eichendorf W., 1984, A\&A, 133, 169 \\
Kwan J. Tademaru E., 1988, ApJ, 332, L41 \\
Kwan J. Tademaru E., 1995, ApJ, 454, 382 \\
Lamzin S.A., Bisnovatyi-Kogan G.S., Errico L., Giovannelli F., Katysheva N.A., Rossi C., Vittone A.A. 1996, A\&A, 306, 877Lago M.T.V.T, 1984, MNRAS, 210, 323 \\
Lago M.T.V.T, Penston M.V., 1982, MNRAS, 198, 429 \\
Lubow S., Artimowicz P., 2000, Protostars and Planets IV, 731 \\
Marsh K.A. Mahoney M.J., 1992, ApJ, 395, L115 \\
Mathieu R.D., Adams F.C., Latham D.W., 1991, AJ, 101, 2184 \\
Mitskevich A.S., Natta A., Grinin V.P., 1993, ApJ, 404,751 \\
Muzerolle J., Hartmann L., Calvet N., 1998, AJ, 116, 455 \\
Najita J., Edwards S., Barsi G., and Carr J., 2000, Protostars and Planets IV, 457
Natta A., Giovanardi C., Palla F., 1988, ApJ, 332, 921 \\
N\"{u}rnberger D., Chini R., and Zinnecker H., 1997, A\&A, 324, 1036 \\
Osterbrock D.E., 1989, Astrophysics of Gaseous Nebulae and Active Galactic Nuclei (Mill Valley: Univ. Science Books) \\
Osterbrock D.E., Tran H.D., Veilleux S., 1992, ApJ, 389, 305 \\
Ray T.P., Mundt R., Dyson J.E., Falle S.A.E.G., Raga A.C., 1996, ApJ, 468, L103 \\
Reipurth B., Pedrosa A., Lago M.T.V.T., 1996, A\&AS, 120, 229 \\
Roddier C., Roddier F., Northcott M.J., Graves J.E., Jim K. 1996, ApJ, 463, 326 \\
Silber J.M., Gledhill T.M., Duch\^ene G., M\'enard F. 2000, ApJL, 536, 89 \\
Solf J., B\"ohm K.H., 1993, ApJ, 410, L31 \\
Takeuchi T., Miyama S.M., Lin D.N.C., 1996, ApJ, 460, 832 \\
\pagebreak

\onecolumn

\begin{table}
\caption{The equivalent widths of the emission lines.} \vspace{0.1cm}
\label{tab:table}
\leavevmode \hspace{3.1cm}
\begin{tabular}[h]{llccc} \hline
Wavelength$^a$ & Line & \multicolumn{3}{c}{Equivalent Width ({\AA})$^b$}\\
\multicolumn{1}{c}{({\AA})} & & 25-Aug-96 & 27-Jun-97 & 2-Jul-99 \\ \hline
6191.56  & FeI          & - & - & 0.057 \\
6238.38  & FeII		& - & - & 0.28 \\
6247.56  & FeII		& - & - & 0.61 \\
6300.23  & [OI]		& - & - & 1.70 \\
6347.09  & SiII		& - & - & 0.23 \\
6363.88  & [OI]		& - & - & 0.43\\
6371.36  & SiII		& - & - & 0.31 \\
6416.91  & FeII		& - & - & 0.27 \\
6432.65  & FeII		& - & - & 0.73 \\
6456.38  & FeII		& - & - & 1.09 \\
6517.02  & FeII         & - & - & 1.08 \\
6562.82  & H$\alpha$	& 145 & 137 & 122 \\
6678.15  & HeI		& 2.24 & 2.09 & 1.23 \\
6583.41  & [NII]	& -$^c$ & -$^c$ & -$^c$  \\
6707.82  & Li           & $-0.16^d$ & $-0.20^d$ & $-0.27^d$ \\
6716.47  & [SII]	& 0.25 & 0.25 & 0.31 \\
6730.85  & [SII]	& 0.50 & 0.57 & 0.73 \\ \hline
\end{tabular} \\ \\
$^a$ References -- Lago \& Penston (1982), Osterbrock, Tran, \& Veilleux (1992)\\
$^b$ Uncertainty of the measurement is 0.1 {\AA} for H$\alpha$, and 0.03-0.05 {\AA}
for the others.\\
$^c$ The equivalent width was not measured because of the difficulty of the removal of the strong H$\alpha$ wing.\\
$^d$ The equivalent widths of the absorption are displayed.\\
\end{table}

\begin{table}
\caption{[SII] equivalent widths \& ratios} \vspace{0.1cm}
\label{tab:table}
\leavevmode \hspace{2.6cm}
\begin{tabular}{ccccc} \hline
Line         & Component      &\multicolumn{3}{c}{Equivalent Width ({\AA})}\\
 & & 25 Aug 96 & 27 Jun 97 & 2 Jul 99 \\ \hline
$[$SII$]$ 6716 {\AA} & HVC          & 0.15 & 0.13 & 0.18 \\
             & LVC          & 0.10 & 0.12 & 0.14 \\
$[$SII$]$ 6731 {\AA} & HVC          & 0.25 & 0.24 & 0.34 \\
             & LVC          & 0.25 & 0.31 & 0.38 \\
6716/6731    & HVC          & 0.60 & 0.54 & 0.53 \\
 ratio       & LVC          & 0.40 & 0.39 & 0.37 \\
\hline
\end{tabular} \\ \\
Note: the dividing velocity for the high velocity component (HVC) and
low velocity component (LVC) is defined as $-95$ km s$^{-1}$ in
the stellar rest frame, as that defined by Hamann (1994).
Uncertainty of
the measured equivalent widths is about 0.02 {\AA}.\\
\end{table}

\pagebreak
\normalsize
\onecolumn


\begin{figure*}
    \leavevmode
\setlength{\unitlength}{1cm}
\begin{picture}(12,12)
\put(8.3,0.5){\psfig{file=takami_Fig1l.epsi,height=11cm}}
\put(-1.1,0.5){\psfig{file=takami_Fig1r.epsi,height=11cm}}
\end{picture}
\caption{Intensity and position spectra obtained on 1999 Jul 2:
(upper left) intensity spectrum at 6160-6480 {\AA},
(upper right) intensity spectrum at 6480-6780 {\AA},
(lower left) position spectrum for the declination at 6160-6480 {\AA}, and
(lower right) position spectrum for the declination at 6480-6780 {\AA}.
}
\end{figure*}

\pagebreak

\begin{figure*}
    \leavevmode
\psfig{file=takami_Fig2.epsi,clip=,width=14 cm}
\caption{The positional displacement of H$\alpha$,
[OI] 6300 {\AA}, [SII] 6731 {\AA}, and [SII] 6716 {\AA} in two-dimensional space
observed on 1999 July 2.
In each map, adjacent velocity components are connected with
solid lines to show the velocity field. Open circles, filled
circles and small crosses in each map corresponds to the velocity
component shown in the intensity profile.
A large cross in each figure shows the position of the continuum source.
Velocity components of each emission lines are plotted
with much less intervals than the spectral resolution
to clarify the spatial structure.
}
\end{figure*}

\pagebreak

\begin{figure*}
    \leavevmode \hspace{4cm}
\psfig{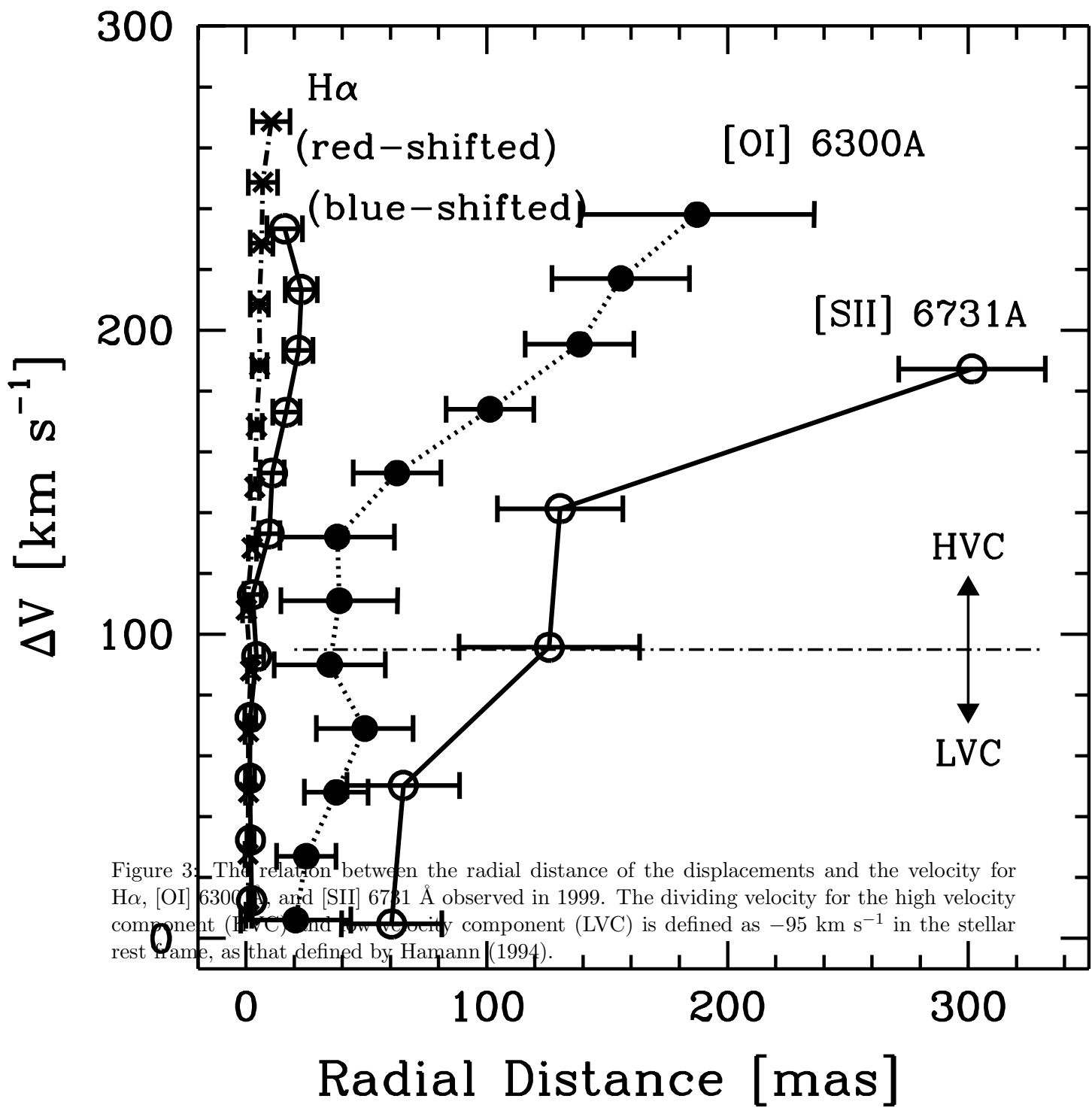}
\caption{The relation between the radial distance of the
                 displacements and
                 the velocity for H$\alpha$, [OI] 6300 {\AA}, and
                 [SII] 6731 {\AA} observed in 1999.
                 The dividing velocity for the high velocity component (HVC) and
                 low velocity component (LVC) is defined as $-95$ km s$^{-1}$ in
                 the stellar rest frame, as that defined by Hamann (1994).
}
\end{figure*}

\pagebreak

\begin{figure*}
    \leavevmode
\psfig{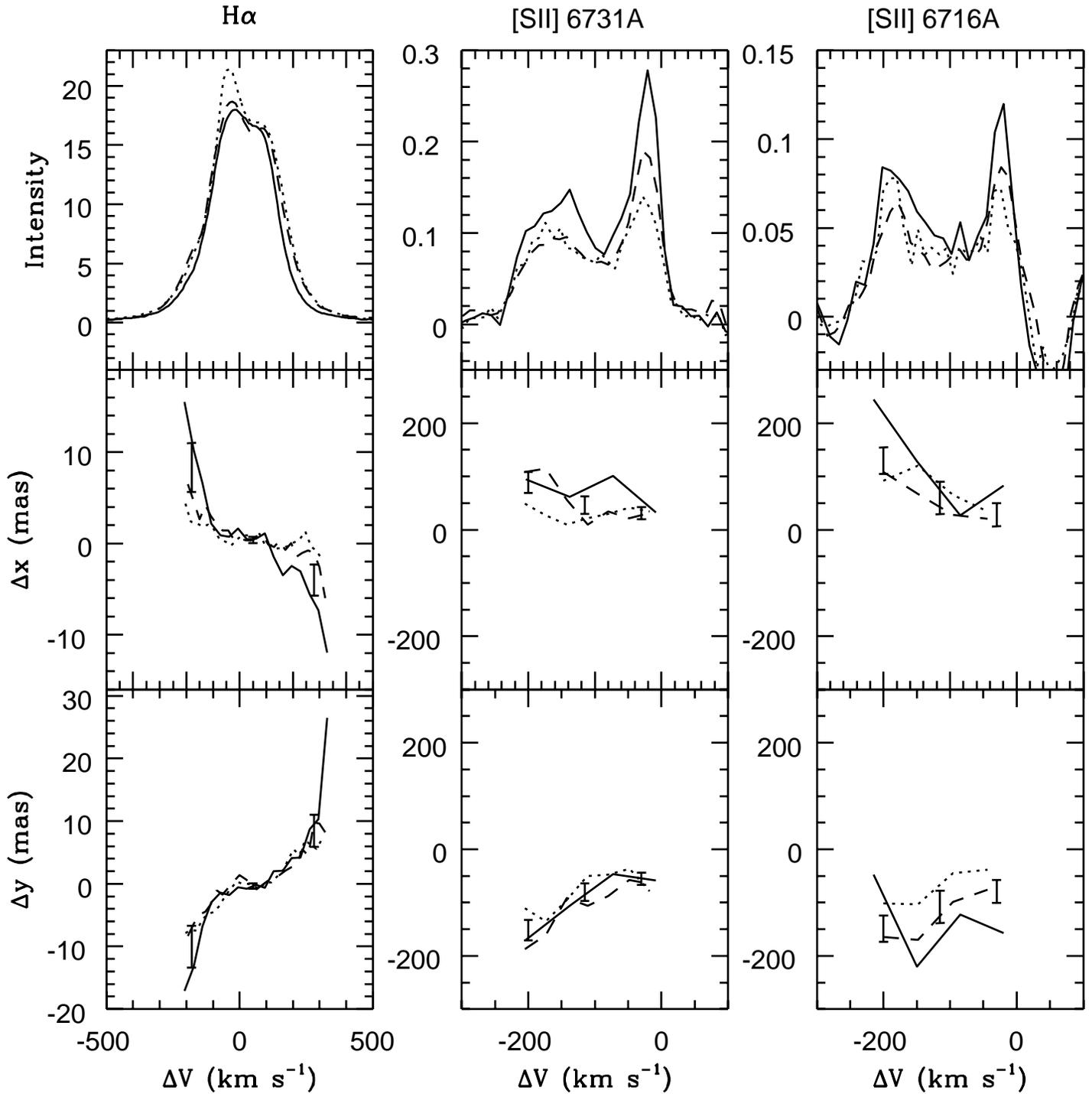}
\caption{ Time variation of the intensity and the positional
                 displacement of H$\alpha$ and [SII] 6731/6716{\AA}
                 emission. Dotted, dashed , and solid lines show
                 the results observed in 1996, 1997, and 1999,
                 respectively. The intensity of each line is
                 normalized by that of the continuum.
                 Vertical bars in the position spectra show
                 typical uncertainties of the measurement. 
}
\end{figure*}

\pagebreak

\begin{figure*}
    \leavevmode
\psfig{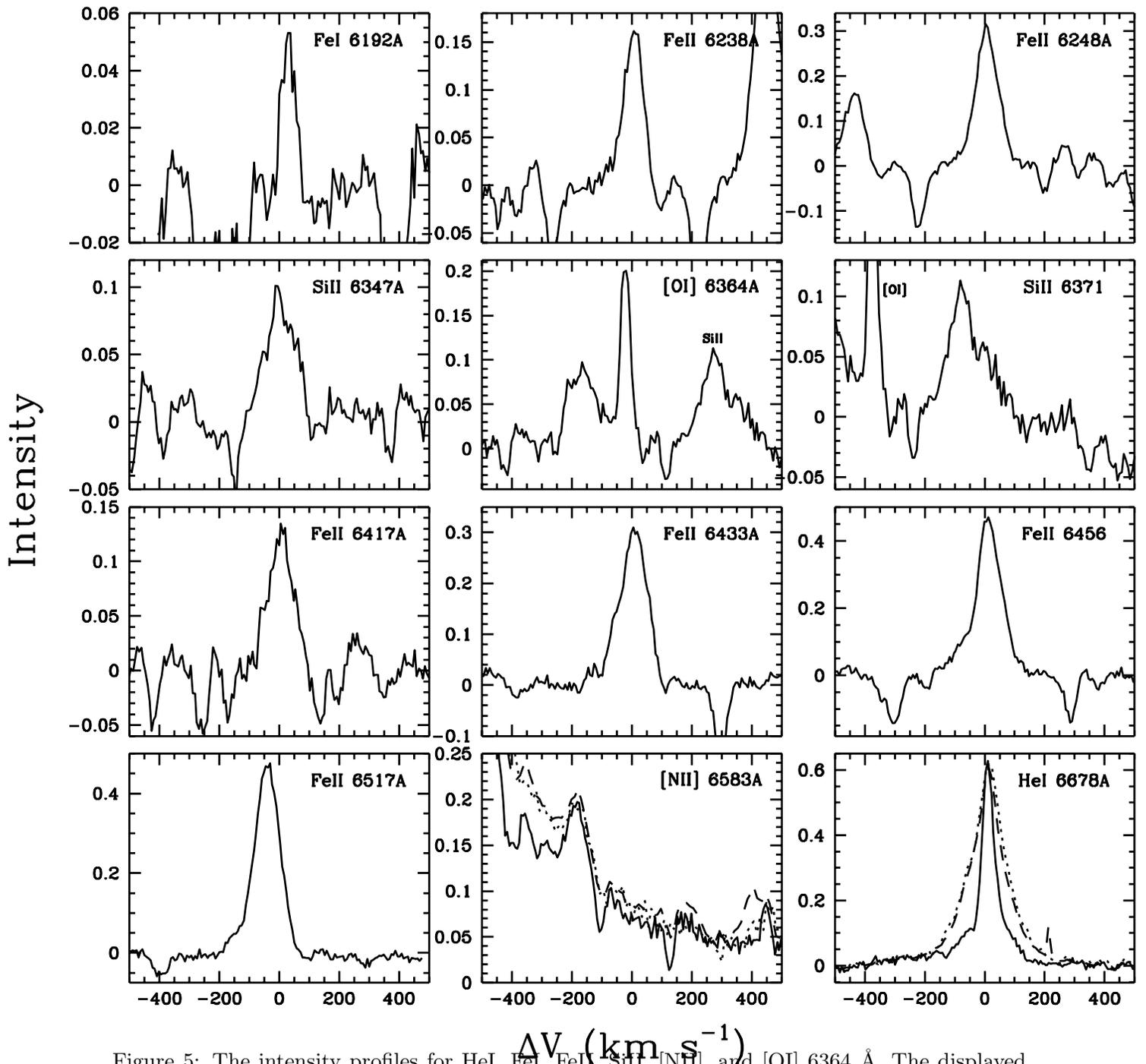}
\caption{The intensity profiles for HeI, FeI, FeII, SiII, [NII], and [OI]
                 6364 {\AA}. The displayed intensity in each figure is normalized
                 by that of the continuum emission at the same wavelength. 
                 Solid curves in all the figures were observed in 1999,
                 while dotted and dashed curves for [NII] and HeI
                 were obtained in 1996 and 1997, respectively.
                 The profiles of [NII] 6583 {\AA}
                 are contaminated by the strong H$\alpha$ wing.
}
\end{figure*}

\pagebreak

\begin{figure*}
\leavevmode \hspace{4cm}
\psfig{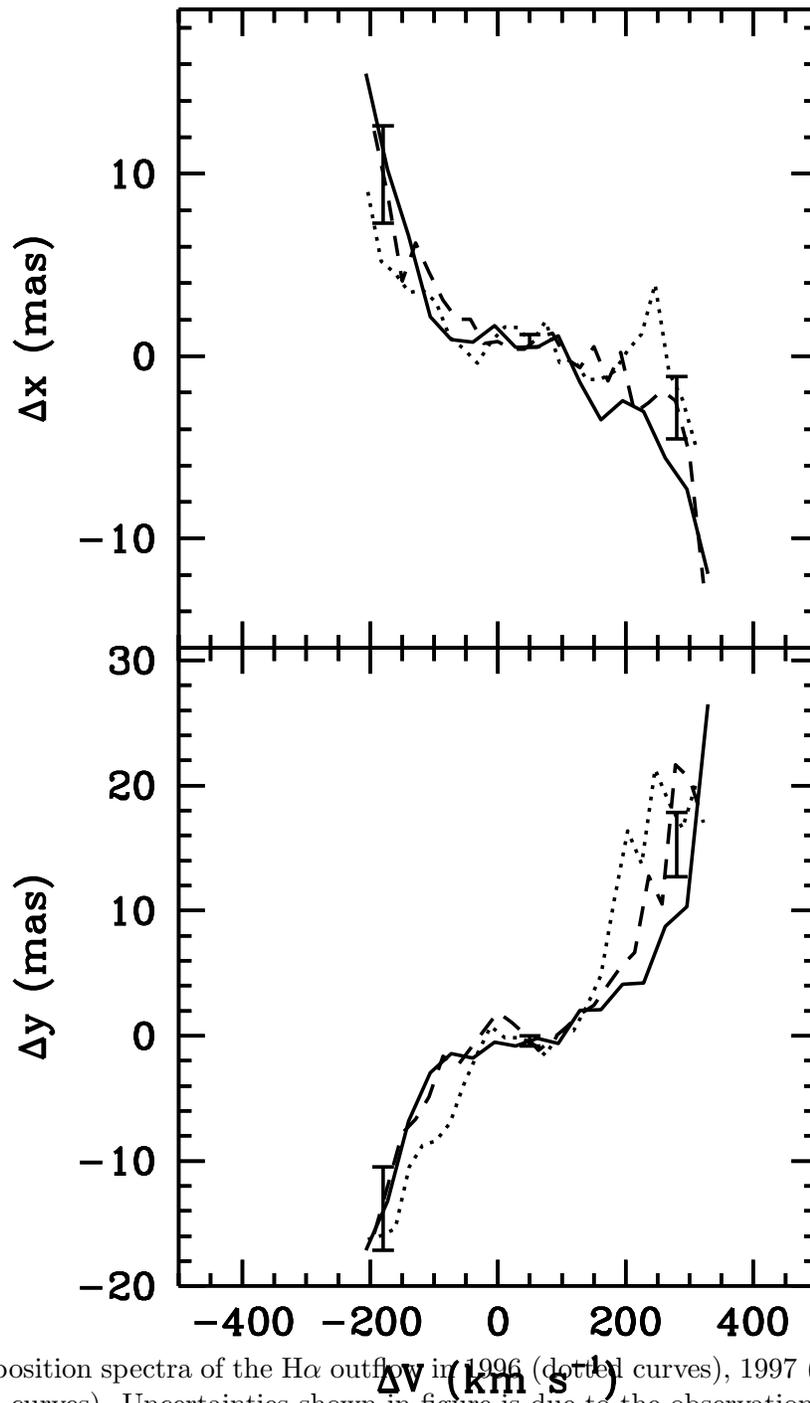}
\caption{The position spectra of the H$\alpha$ outflow
                 in 1996 (dotted curves),
                 1997 (dashed curves), and 1999 (solid curves).
                 Uncertainties shown in figure is due to
                 the observations.
}
\end{figure*}

\pagebreak

\begin{figure*}
    \leavevmode \hspace{4cm}
\psfig{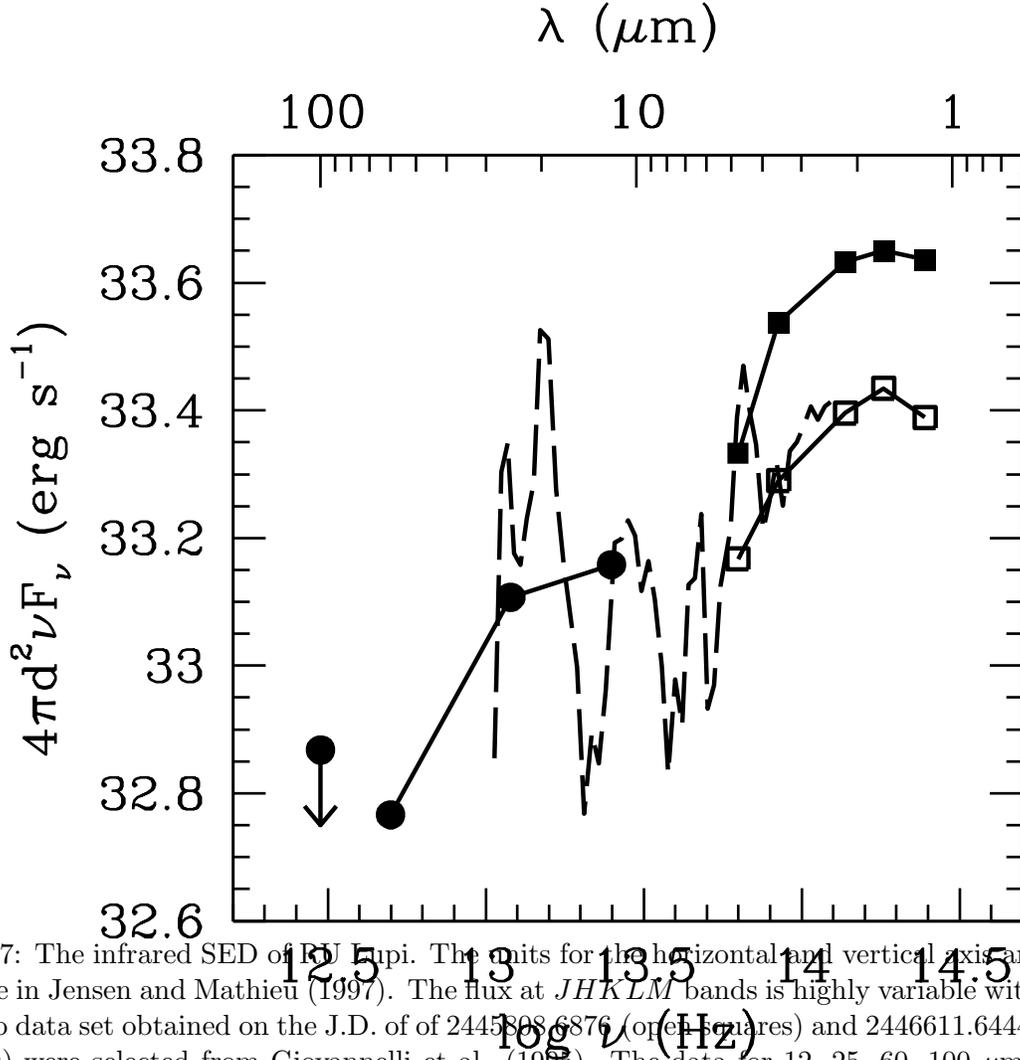}
\caption{The infrared SED of RU Lupi. 
The units for the horizontal
and vertical axis are same as those in Jensen and Mathieu (1997).
The flux at $JHKLM$ bands is
highly variable with time, and two data set obtained on the J.D. of
of 2445808.6876 (open squares) and 2446611.6444 (filled squares) were selected
from Giovannelli et al. (1995).
The data for 12, 25, 60, 100 $\mu$m shown with filled circles were obtained
by $IRAS$ observations. A mid-infrared spectrum obtained
by ISO-SWS was smoothed along the wavelength to achieve higher signal-to-noise
ratio, and plotted with a dashed line. The resultant spectral resolution
($\lambda / \delta \lambda$) of the spectrum is 20.
}
\end{figure*}

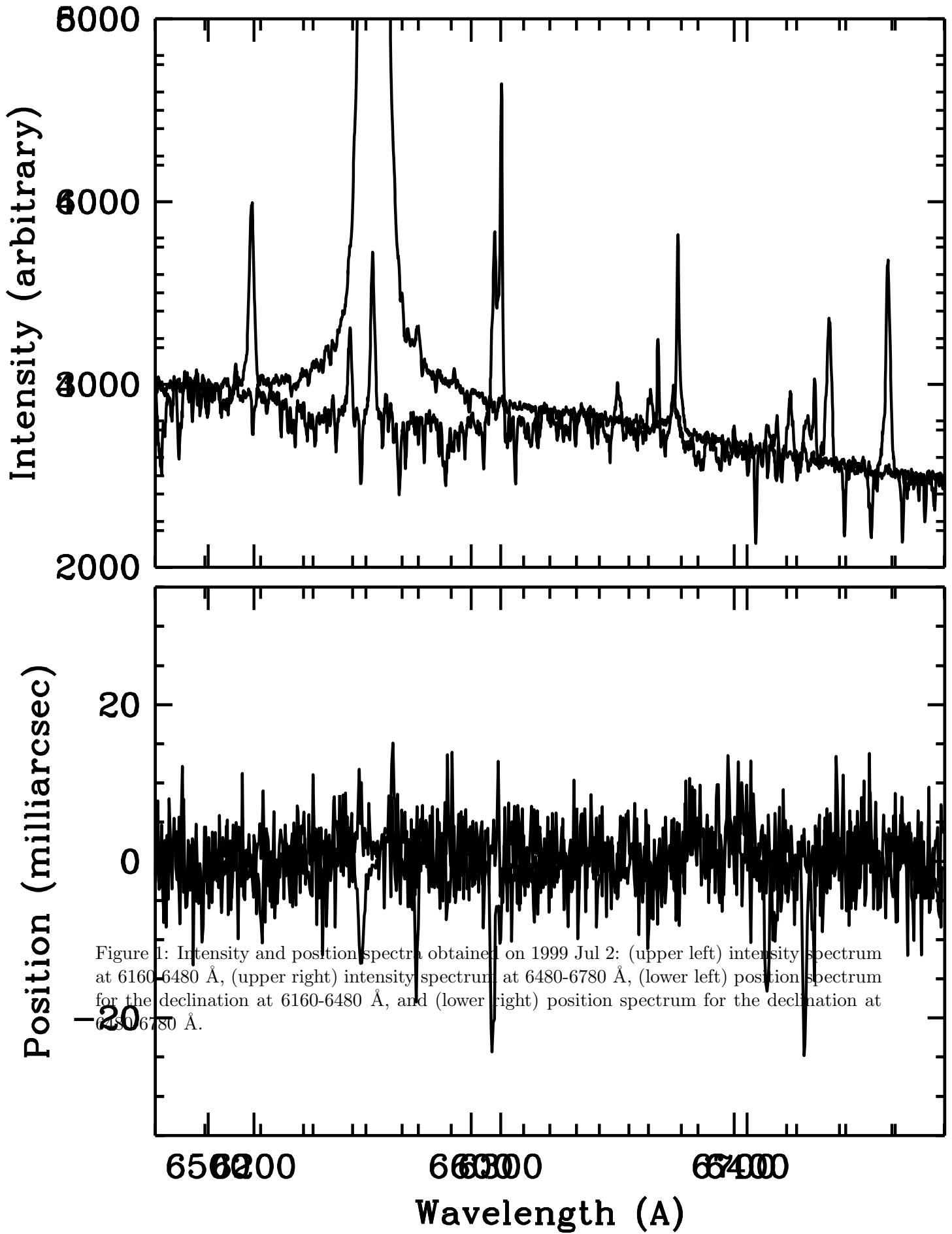
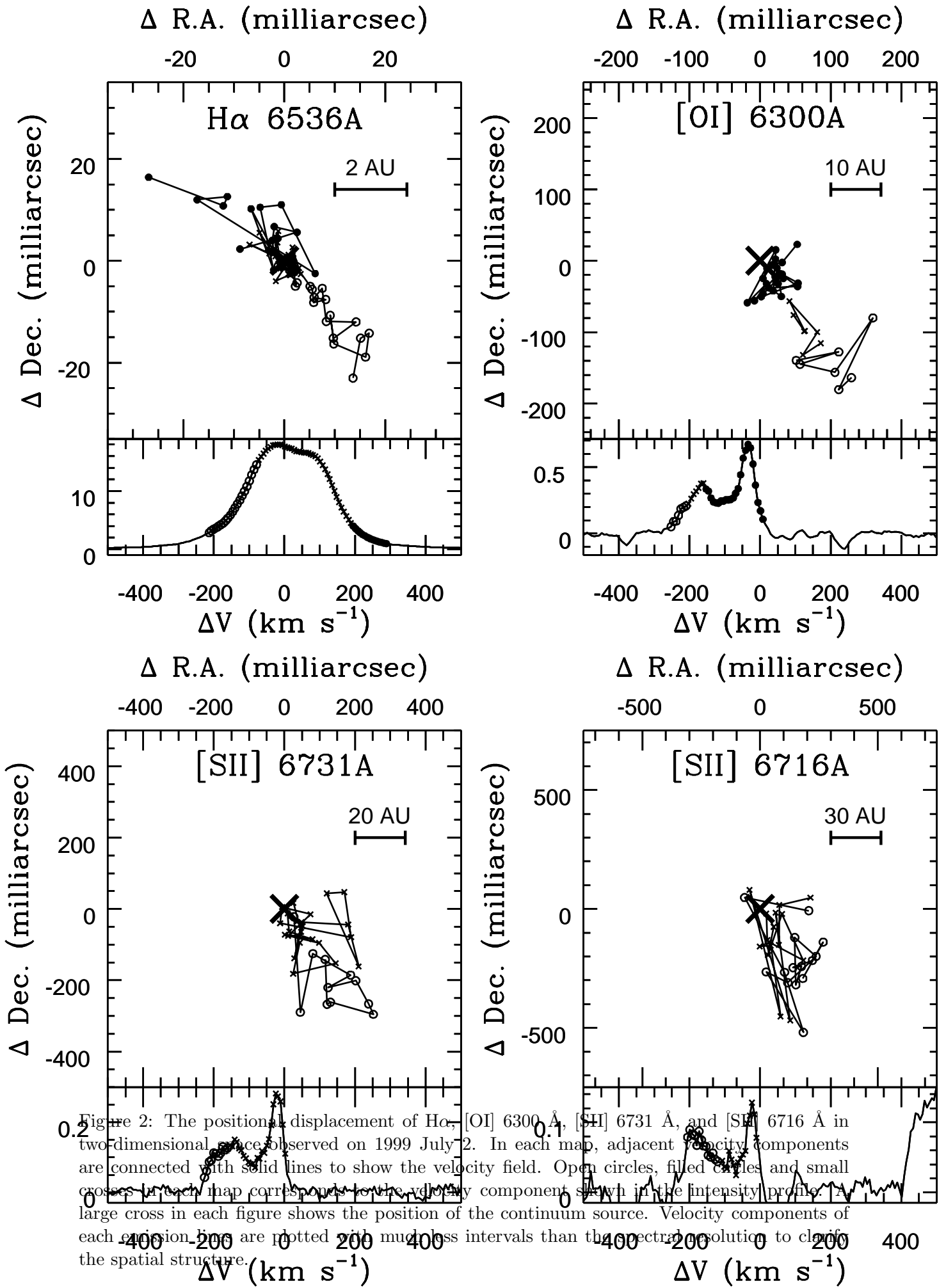
\end{document}